\documentclass[article]{elsarticle}

\makeatletter
\def\ps@pprintTitle{
  \let\@oddhead\@empty
  \let\@evenhead\@empty
  \let\@oddfoot\@empty
  \let\@evenfoot\@oddfoot
}

\makeatother\usepackage{amsmath}

\begin{document}

\begin{frontmatter}

\title{Extending the Lorentz Factor Range and Sensitivity of Transition Radiation with Compound Radiators}

\renewcommand*{\thefootnote}{\fnsymbol{footnote}}

\author{Samer T. Alnussirat\footnote[1]{Current address: Space Sciences Laboratory, Univ. of Calif., Berkeley, CA 94720, USA}  
 \& Michael L. Cherry\footnote[2]{Corresponding author: Michael L Cherry, cherry@lsu.edu}
 }

\address{Dept. of Physics $\&$ Astronomy,  Louisiana State University, Baton Rouge, LA 70803 USA}

\begin{abstract}
Transition radiation detectors (TRDs) have been used to identify high-energy particles (in particular, to separate electrons from heavier particles) in accelerator experiments. In space, they have been used to identify cosmic-ray electrons and measure the energies of cosmic-ray nuclei. To date, radiators have consisted of regular configurations of foils with fixed values of foil thickness and spacing (or foam or fiber radiators with comparable average dimensions) that have operated over a relatively restricted range of Lorentz factors. In order to extend the applicability of future TRDs (for example, to identify 0.5 - 3 TeV pions, kaons, and protons in the far forward region in a future accelerator experiment or to measure the energy spectrum of cosmic-ray nuclei up to 20 TeV/nucleon or higher), there is a need to increase the signal strength and extend the range of Lorentz factors that can be measured in a single detector. A possible approach is to utilize compound radiators consisting of varying  radiator parameters. We discuss the case of a compound radiator and derive the yield produced in a TRD with an arbitrary configuration of foil thicknesses and spacings.

\end{abstract}


\end{frontmatter}


\section{Introduction}

Transition radiation is a classical electromagnetic phenomenon closely related to bremsstrahlung and Cerenkov radiation. First proposed theoretically by Ginzburg and Frank  \cite{ginzburg1946radiation} in 1946, its existence was soon demonstrated experimentally at Yerevan and Brookhaven  \cite{alikhanian1973detection,yuan1970energy}. The TREE cosmic-ray balloon experiment \cite{prince1979energy} demonstrated the capability of a transition radiation detector (TRD) in combination with a thin calorimeter to identify electrons in the presence of a flux of protons more than 2 orders of magnitude larger, and since then TRDs have been used in a number of cosmic-ray experiments up to the AMS experiment \cite{kirn2004ams} currently flying on the International Space Station. Most of these cosmic-ray applications have been aimed at identifying electrons, but the Spacelab II \cite{grunsfeld1988energy} and TRACER \cite{ave2008composition} instruments extended the use of TRDs to measure the energy spectra of high-energy cosmic-ray nuclei. At accelerators, most applications have used TRDs to identify electrons (as examples, see refs. \cite{boldyrev2012atlas, Busch:2013ica}), and the ATLAS group has recently conducted an extensive series of beam exposures that have enabled detailed comparisons of the experimental results with GEANT predictions \cite{alozy2019identification, savchenko2020fine, alozy2020studies}.

As a charged particle with Lorentz factor $\gamma \gg$ 1 crosses the boundary between two materials of different dielectric constant, a homogeneous solution to Maxwell’s equations must be introduced in order to match the boundary conditions. This added field component due to the presence of the interface results in radiation at X-ray frequencies extending up to $\omega \sim  \gamma \; \omega_1$, where (in the case of the interface between a solid and vacuum) $\omega_1$ is the plasma frequency of the solid material. The total emitted intensity in the case of a single interface increases linearly with $\gamma$.

\begin{figure}[htbp]
\begin{center}
\includegraphics[width=1\textwidth]{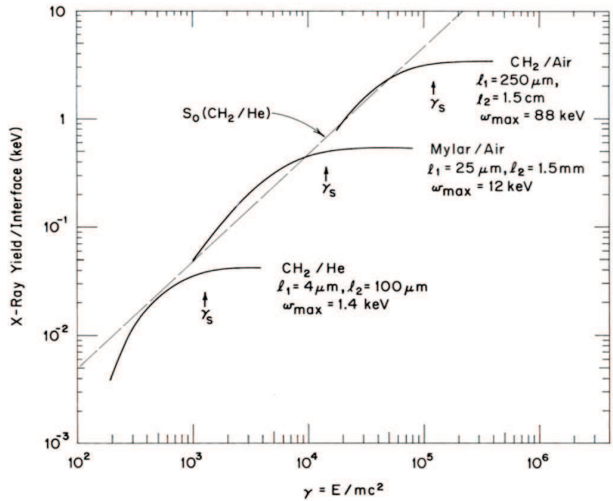}
	\caption{\label{fig1}Total emitted intensity radiated per interface for a single CH$_2$/He interface (dashed line) and three multi-foil radiator configurations, demonstrating ability to tune radiator parameters \cite{cherry1974transition}.}	
\end{center}
\end{figure}

Unfortunately, the radiation intensity is low: The probability of producing a photon is $\sim 10^{-2}$/ interface. To build up a detectable signal, one typically allows the particle to pass through a periodic stack of several hundred or thousand foils of fixed thickness each separated by a fixed spacing of gas or vacuum \cite{ter1961emission,G14}. If the field amplitudes at each interface are added in phase, it can be shown that the result is an interference pattern, with peaks appearing at frequencies governed by a resonance condition determined by the foil thickness $l_1$ and spacing $l_2$ \cite{cherry1974transition, artru1975practical, cherry1978measurements, dolgoshein1993transition}. As $\gamma$ increases, the spectrum extends to successively higher frequencies, with the largest contribution to the emitted energy appearing at the highest frequency maximum near $\omega_{max} \sim \omega_1^2 l_1/2\pi c$. Once this highest maximum appears, there is no additional contribution to the intensity and the emission saturates near a Lorentz factor $\gamma_s \sim 0.6\omega_1 \sqrt
{l_1 l_2}/c$ \cite{cherry1974transition,artru1975practical, cherry1978measurements}. The dependence of $\omega_{max}$ and $\gamma_s$ on the radiator parameters $l_1$ and $l_2$ means that the performance can be tuned to fit the application (Fig. \ref{fig1}).

The emitted X-rays are produced in a narrow range of angles $\theta \sim 1/\gamma$ in the forward direction, so that (in the absence of a bending magnet to deflect the particle), the X-rays are detected together with the particle ionization. This typically implies either that the detector must be thin (in order to maximize the ratio of detected X-ray signal to $dE/dx$) and/or the number of foils must be large (although not so large as to produce interactions, delta rays, or bremsstrahlung background). Designing a TRD for an application therefore involves matching the radiator materials, foil thickness, and spacing to the desired range of Lorentz factors, and matching the frequency spectrum to the detector efficiency and radiator transmission.  

In Fig. 1, the number of X-ray photons produced in a typical radiator of 100 foils is very roughly the yield in keV/interface times 200 interfaces divided by $\hbar\omega_{max}$. It can be seen from Fig. 1 that the range in Lorentz factor over which the emitted radiation increases from approximately a single emitted X-ray to saturation is typically no more than a factor $\sim$ 5. There are currently new experiments being proposed that will require a wider operating range than this. As examples: 1) The Forward Multiparticle Spectrometer (FMS) project \cite{albrow2020forward,C20} currently being discussed at CERN may require a TRD to identify 0.5 - 3 TeV pions, kaons, and protons; and 2) the Advanced Particle-astrophysics Telescope (APT) \cite{buckleyICRC} is a mission concept for a future space-based gamma-ray telescope that may also have the capability to measure the spectrum of cosmic-ray Boron and Carbon at 500 GeV/nucleon up to Iron at 20 TeV/nucleon. In both cases, an appropriate TRD  must operate over the $\gamma$ range from $\sim 500$ to at least  $2\times 10^4$. 

Current TRDs are not able to operate over such a wide range of $\gamma$ values. One possibility is to employ multiple TRDs in sequence - for example a standard periodic “regular” radiator with a “high-energy” segment with large fixed values of  $l_1$ and $l_2$ followed by a “low-energy” segment with small  $l_1$ and $l_2$ \cite{Belyaev,Wakely}. A second possibility is to design a “compound” radiator with a customized arrangement of multiple $l_1$ and $l_2$ values to extend the energy range to values both lower and higher than typically covered by current TRDs.
General considerations of the radiation produced in a compound (irregular) radiator have been presented previously \cite{Grichine, GaribyanIrregular}. In this paper, we present a relatively simple derivation of the radiation intensity produced in a compound radiator and emphasize the potential to extend the $\gamma$ range of the detector in a single device by using compound rather than regular radiators. 

In Sec. II, we review the expressions for the signal produced in radiators with regular fixed values of foil thickness $l_1$ and gap spacing $l_2$. In Sec. III, we then calculate the signal produced by compound radiators consisting of foils with arbitrary values of thickness and spacing designed for both low and high energies. We present numerical examples in Sec. IV, and conclude in Sec. V.

\section{Radiators with Fixed Values of Foil Thickness and Gap Spacing}
In a regular radiator consisting of $N$ identical foils of thickness $l_1$ separated by equal distances $l_2$, the field amplitudes at each interface must be added in phase. The resulting intensity per unit frequency $\omega$ per unit solid angle $\Omega$ is given by the expression \cite{ter1961emission, G14,cherry1974transition, artru1975practical,cherry1978measurements, dolgoshein1993transition}

\begin{equation}
\frac{d^2S_N}{d\Omega d\omega } = \frac{d^2S_0}{d\Omega d\omega } 4 \sin^2\frac{l_1}{Z_1} \frac{\sin^2 N\Big(l_1/Z_1 +l_2/Z_2\Big)}{\sin^2\Big(l_1/Z_1 +l_2/Z_2\Big)} \; \; .
\end{equation}
Here
\begin{equation}
\frac{d^2S_0}{d\Omega d\omega }=\frac{1}{c}\Big(\frac{qe\omega\theta}{4\pi c}\Big)^2 (Z_1 - Z_2 )^2
\end{equation}
is the intensity emitted at angle $\theta$ with respect to the beam by a particle of charge $qe$ , velocity $\beta = v/c$, and Lorentz factor $\gamma \gg 1$ traversing a single interface between medium 1 and medium 2 (typically a solid foil and a gas or vacuum gap). The plasma frequencies in the two media are $\omega_1$ and $\omega_2$, where $\omega_{1,2} \ll \omega$, and the quantities
\begin{equation}
Z_{1,2} = \frac{4 \pi}{c}\Bigg(\frac{1}{\gamma^2} +  \frac{\omega_{1,2}^2}{\omega^2} +\theta^2\Bigg)^{-1}
\end{equation}
are known as the formation zones in the two materials.

The frequency spectrum from a periodic regular radiator of identical foils extends up to a characteristic X-ray frequency $\gamma \omega_1$, where $Z_1-Z_2$ approaches 0 in Eq. 2. At every frequency $\omega$, the factor
$\sin^2 N(l_1/Z_1+l_2/Z_2) \: /\: \sin^2 (l_1/Z_1+l_2/Z_2) $ acts like a delta function $N \pi \delta(l_1/Z_1 + l_2/Z_2 - \pi r)$ during the integration over angles, leading to emission at angles
\begin{equation}
\theta^2 = \frac{4 \pi c}{\omega} \; \frac{r-r_{min}}{l_1+l_2 },
\end{equation}
where $r$ is an integer
\begin{equation}
r > r_{min} = \frac{\omega}{4 \pi c}\Bigg[l_1 \Bigg(\frac{1}{\gamma^2}+\frac{\omega_1^2}{\omega^2} \Bigg)+l_2 \Bigg(\frac{1}{\gamma^2}+\frac{\omega_2^2}{\omega^2}\Bigg) \Bigg] \; \; .
\end{equation}

The frequency spectrum integrated over angles (i.e., summed over r)  displays an interference pattern. In practice, photoelectric attenuation in the radiator typically attenuates the low frequency signal, and (for sufficiently energetic incoming particles) the largest signal typically appears in the highest frequency maximum near $\omega_{max}$. As the particle energy increases, the spectrum hardens and the  total emitted energy increases roughly linearly with $\gamma$ until the highest frequency maximum appears and the signal saturates  near Lorentz factor $\gamma_s$.

\section{Compound Radiators with Arbitrary Foil Dimensions}
Consider a simple example of a radiator consisting of $N$ compound foils, each consisting of two sub-foils with respective thicknesses $l_1$ and $l_2$, with a gap of thickness $l_0$ between foils. The plasma frequencies are $\omega_i$, where $i = 0, 1, \text{or}\;  2$. The total electric field is computed by summing the field amplitudes at each interface between medium $i$ and medium $j$ in each foil $n$:
\begin{equation}
E = \sum_{n=1}^{N}\Big[\frac{A_{01}}{R}e^{i \phi_{01}^n} + \frac{A_{12}}{R}e^{i \phi_{12}^n} +\frac{A_{20}}{R}e^{i \phi_{20}^n} \Big]    .
\end{equation}
The amplitudes $A_{ij}$ corresponding to the transition from medium i to medium j are given by \cite{cherry1974transition,Garibian}
\begin{equation}
A_{ij} = \frac{q e \beta \omega \sin\theta \cos\theta}{2 \pi c^2}\big(Z_i-Z_j\big)    .
\end{equation}
Here $R$ is the distance to the observation point and $\theta$ is the angle of emission. 

At high energies $ \gamma \gg 1$ and $\omega \gg \omega_i$, the phase factor $\phi_{ij}^n = \vec{k} \cdot \vec{x} - \omega t$ at the interface between media $i$ and $j$ in each foil $n$ can be written in terms of the dielectric constants $\epsilon_i = 1-w_i^2/\omega^2$ as

 \begin{equation} 
\begin{split}
& \phi_{12}^n  = \phi_{20}^n+\frac{\omega}{c} \Big(l_2 \sqrt\epsilon_2\cos\theta +\frac{l_2}{\beta}\Big) = \phi_{20}^n-2\frac{l_2}{Z_2},\\
& \phi_{01}^n  = \phi_{20}^n-2\Big(\frac{l_1}{Z_1}+\frac{l_2}{Z_2}\Big),\\
& \phi_{ij}^n-\phi_{ij}^1  =-(n-1)\Delta\phi,\\
& \Delta\phi  =2\Big(\frac{l_0}{Z_0}+\frac{l_1}{Z_1}+\frac{l_2}{Z_2}\Big)\; \; .
\end{split}
\end{equation}
 
 To within an overall phase factor, the summed electric field amplitude then becomes

   \begin{equation}
E = \Big[\frac{A_{01}}{R}e^{-2i(l_1/Z_1+l_2/Z_2)} + \frac{A_{12}}{R}e^{-2il_2/Z_2} +\frac{A_{20}}{R} \Big]\frac{\sin N\Delta\phi/2}{\sin \Delta\phi/2} \; \; .
 \end{equation}
 
 \noindent  The intensity radiated per unit angle $\theta$ and frequency $\omega$ can be written in terms of the Poynting vector flux 
\begin{equation}
\frac{d^2S_N}{d\Omega d\omega } = \frac{c}{4\pi} E H^*R^2 = \frac{c}{4 \pi} \chi^2\frac{\sin^2 N\Delta\phi/2}{\sin^2 \Delta\phi/2}
\end{equation}
 with
\begin{equation} 
\begin{aligned}
 \chi^2 = {}& A_{01}^2+A_{12}^2+A_{20}^2  \\
 &+2A_{01}A_{12}\cos(2l_1/Z_1) \\
 &+2A_{12}A_{20}\cos(2l_2/Z_2)  \\
 &+2A_{20}A_{01}\cos2(l_1/Z_1+l_2/Z_2)  . 
 \end{aligned}
 \end{equation}
 In the case where the compound foil reduces to a simple foil (i.e, medium 2 disappears), Eq. 10 reduces to the regular foil formula (1).
 
 For $N$ large, Eq. 10 can be integrated over angles by using the fact that $\frac{\sin^2 N\Delta\phi/2}{\sin^2 \Delta\phi/2} $ again behaves like a delta function. For each frequency $\omega$, the resulting summation therefore picks out the angles $\theta$ satisfying the condition 
 
\begin{equation}
\frac{l_0}{Z_0}+\frac{l_1}{Z_1}+\frac{l_2}{Z_2} = \pi r
\end{equation} 
for $r > r_{min}$, where (for a radiator with $N$ foils, each with $M$=2 sub-foils)
\begin{equation}
\begin{aligned}
\theta^2 &= \frac{4 \pi c}{\omega} \; \frac{r-r_{min}}{\sum_{m=0}^{M} l_m}\\
r_{min} &=  \frac{\omega }{4 \pi c}\sum_{m=0}^{M}l_m  \Big(\frac{1}{\gamma^2}+\frac{\omega_m^2}{\omega^2}\Big)  \; \; .
\end{aligned}
\end{equation} 

The resulting frequency spectrum is
\begin{equation}
\frac{dS}{d\omega} = \frac{N q^2 e^2\omega}{4\pi c^2 {\sum_{m=0}^{M}} \; l_m} {\sum_{\; \; r=r_{min}}^{\infty}} \theta^2 \Lambda_M
\end{equation}
where
\begin{equation} 
\begin{aligned}
\Lambda_2 = {}& (Z_0-Z_1)^2+(Z_1-Z_2)^2+(Z_2-Z_0)^2  \\
 &+2(Z_0-Z_1)(Z_1-Z_2)\cos(2l_1/Z_1) \\
 &+2(Z_1-Z_2)(Z_2-Z_0)\cos(2l_2/Z_2)  \\
 &+2(Z_2-Z_0)(Z_0-Z_1)\cos2(l_1/Z_1+l_2/Z_2)  \; \; .
\end{aligned}
\end{equation}

Formulas 13-15 are derived for the special case of $N$ foils each consisting of $M$=2 sub-foils. To generalize to the more general case of $M>2$ sub-foils, we assume that medium $M+1$ is a repeat of medium 0;  i.e. $\omega_{M+1}=\omega_0, \; l_{M+1}=l_0, \; Z_{M+1}=Z_0,$. It can then be shown that
\begin{equation} 
\begin{aligned}
\Lambda_M ={} & \sum_{m=0}^{M}(Z_m-Z_{m+1})^2 
+2\sum_{m=0}^{M-1}(Z_m-Z_{m+1})  \\
&\times \sum_{n=m+1}^{M}(Z_n-Z_{n+1})\cos \Big(2\sum_{i=m+1}^{n} l_i/Z_i\Big) \; \; .
\end{aligned}
\end{equation}
The configuration of N foils each consisting of M sub-foils with arbitrary plasma frequencies and dimensions  is equivalent to the general case of N separate modules each consisting of M individual foils. 

\section{Numerical Example}
As discussed above, the accelerator and cosmic ray applications FMS and APT have similar requirements for the $\gamma$ range where the TRD must operate. Belyaev et al. \citep{Belyaev} have proposed using a combination of periodic “high-energy” and “low-energy” radiators to perform the $\pi$-K-p identification needed for FMS. Here we use the APT example to describe the design and potential performance of a suitable compound radiator.

The cosmic-ray Boron-to-Carbon (secondary-to-primary) ratio has been measured up to $\sim$ 2 GeV/nucleon and the spectra of Carbon through Iron nuclei have been measured up to approximately 2 TeV/nucleon \cite{CREAM, TRACER, NUCLEON, aguilar2017observation, aguilar2018observation,adriani2020direct,adriani2021measurement}. In order to measure the Boron-to-Carbon ratio near 500 GeV/nucleon and the spectra of cosmic-ray nuclei up to 20 TeV/nucleon on a space instrument with limited mass, where a heavy calorimeter or magnetic spectrometer is probably not feasible, a TRD with sensitivity covering the Lorentz factor range $500 <\gamma< 2 \times 10^4$ may be a viable approach. 

A possible future instrument suitable for the cosmic-ray measurement might be the Advanced Particle-astrophysics Telescope (APT), currently a design study for a future MIDEX or probe-class mission \cite{buckleyICRC}. The instrument is planned as a large (3m $\times$ 3m or 3m $\times$ 6m) space-based gamma-ray/cosmic-ray instrument intended to a) test the thermal WIMP dark matter paradigm and b)  provide prompt localization for electromagnetic counterparts of gravity wave/neutron star mergers. APT is currently designed to have an order of magnitude greater gamma-ray sensitivity than Fermi LAT at GeV energies, simultaneously providing sub-degree MeV transient localization over the largest possible field of view. The very large area needed to achieve this sensitivity coupled with a high Earth orbit dictate the use of a calorimeter with limited depth ($<$ 6  radiation lengths) to reduce mass. The current APT design incorporates 20 layers of 5 mm thick CsI:Na with crossed wavelength-shifting fiber readout, interspersed with 20 scintillating optical fiber tracker layers. Such an instrument could also be a powerful cosmic-ray detector. With the addition of a TRD, the transition radiation X-ray signal from very high energy light cosmic-rays (i.e., Boron and Carbon) could provide information to differentiate models essential for interpreting positron and antiproton data to look for a dark matter signature, and the measurement of the nuclear spectra through Iron could shed light on the cosmic-ray propagation effects that appear to be responsible for the hardening of the nuclear spectra seen at energies beginning near 200 GeV/nucleon (for example, in refs. \cite{CREAM} - \cite{adriani2021measurement}).

The weight of the large-area CsI scintillators is supported by 5 cm of plastic foam below each CsI layer. By replacing the current foam layer with an appropriate set of TR radiators and adding a layer of Xenon gas detectors to detect the TR X-rays, a cosmic-ray capability could be added to APT. In order to cover the Lorentz factor range from 500 to $2 \times 10^4$ and fit into the current 5 cm space available for the radiator, two approaches appear feasible: 
\newline
\textbf{a) Combination of regular “low-energy” and “high-energy” radiators:} If the 5 cm of plastic foam in the current APT design is chosen with cell wall thicknesses and cell diameters corresponding to an appropriate $l_1$ and $l_2$, the foam can potentially be implemented as a TR radiator. A number of competing requirements must be satisfied: In order to operate at $\gamma$ values as low as 500, the TRD must incorporate a “low energy” section with small $l_1$ and $l_2$. As is demonstrated in Fig. 1, the signal per interface from such a radiator is small, and so the number of interfaces (i.e., foam cells) must be large. The typical frequency $\omega_{max}$ is low as well, and so the photoelectric absorption must be minimized (i.e., the foam must consist of low $Z$ material and the total thickness in g/cm$^2$ must be small). In order to provide a useful energy dependence up to $\gamma$ values $\sim 2\times10^4$, a “high energy” section of the radiator must then consist of larger cells with thicker walls. The need to maximize the emitted signal (i.e., increase the number of interfaces N) must again be balanced against the need to minimize the X-ray absorption. Since APT’s prime science goal is to measure the cosmic gamma-ray signal, the total thickness of the radiators must be small enough to have only a small effect on the $\gamma-$ray transmission. Finally, the total radiator thickness must be small enough to produce only a small nuclear interaction probability and $\delta$-ray background.

As an initial example of a “low energy - high energy” radiator combination based on periodically spaced foils, we consider (Configuration a) a regular “high energy” (HE) radiator followed by a regular “low energy” (LE) radiator. The HE radiator is composed of 11 CH$_2$ foils each 50 $\mu$m thick with a vacuum gap between successive foils of thickness 4 mm. This saturates at $\gamma_s \sim 2.8 \times 10^4$, with a peak in the TR spectrum at $\omega_{max} \sim $ 18 keV. The total thickness of this radiator is 4.05 cm. The HE radiator is followed by an LE module consisting of 30 CH$_2$ foils each 25 $\mu$m thick with a vacuum gap between foils of thickness 0.2 mm. This saturates at $\gamma_s \sim$ 4500, with a peak in the TR spectrum at $\omega_{max} \sim $ 9 keV. The total thickness of this radiator is 0.675 cm. It is followed by two layers of 2 mm thick Xe straw tubes to detect the TR signals. The crossed Xe layers replace the current scintillating fiber tracker layers. The total thickness of this composite radiator combination is 5.1 cm, comparable to the thickness of the foam plus scintillating fibers in the current APT design. The total detector consists of 20 layers, each layer consisting  of 5 mm of CsI plus a CH$_2$-vacuum (or CH$_2$-He) radiator plus a pair of Xe straw tubes. As an alternative (Configuration b), in order to increase the signal with minimal change in overall length but at the cost of reduced $\gamma_s$, we replace the HE module with $N$ = 22 CH$_2$ foils with $l_1$ = 50 µm and $l_2$ = 2 mm.

\textbf{b)  Compound radiator:} A second approach incorporates a compound radiator (Configuration c) with 120 foils ($N$ = 20, $M$ = 120 in Eqs. 14 and 16), with the foil thickness decreasing by a constant factor from 100 $\mu$m to 11 $\mu$m and the spacing decreasing steadily from 2800 to 93 $\mu$m; i.e., the radiator parameters decrease from a configuration corresponding to $\gamma_s$ = $3.3 \times 10^4$ and $\omega_{max}$ =35  keV at the entrance of the radiator to $\gamma_s$ = 2000 and $\omega_{max}$ = 3.9 keV at the exit. The total thickness of the compound radiator is  5.1 cm corresponding to 0.2 g/cm$^2$ total grammage, 0.004 radiation lengths, and 0.004 nuclear collision lengths. It should be noted that rather than using a set of foils manufactured with the appropriate $l_1$ and $l_2$, a suitable compound radiator could be produced using a graded sample of hollow plastic microspheres produced by emulsion polymerization and available commercially in suitable dimensions.

\begin{figure}[htbp]
\begin{center}
\includegraphics[width=1\textwidth]{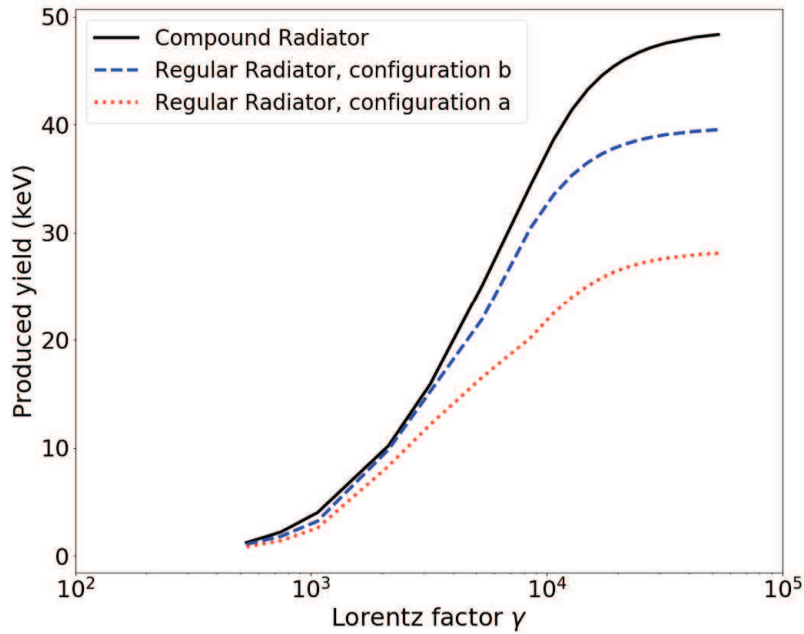}
	\caption{\label{fig2}Emitted transition radiation signal (keV) vs Lorentz factor produced by a singly charged particle passing through a module of compound radiator (Configuration $c$, solid line) and two regular HE-LE radiators (Configuration $a$, dotted line; Configuration $b$, dashed line).}	
\end{center}
\end{figure}

The solid curve in Fig. \ref{fig2} shows the average emitted energy in keV as a function of Lorentz factor produced by an incident electron or proton passing through a single  compound radiator module. The dotted and dashed curves show the signal for an incident particle passing through Configurations a and b. The dependence on energy for these initial “low energy - high energy” configurations indeed show  a wider useful range of Lorentz factors than the typical factor of $\sim$ 5 in Fig. 1. Neither the compound nor the two regular radiator configurations a or b have been subjected to a rigorous optimization analysis, but Fig. 2 demonstrates that for this set of configurations with comparable total length (i.e., designed to fit within the practical constraints of the prospective FMS and APT experiments), a compound radiator produces a higher yield and wider range of Lorentz factor sensitivity than comparable regular radiators. (X-ray absorption is neglected in these initial calculations of feasibility.)

 It should be noted that the higher yield for the compound radiator is due partly to the use of a larger number of foils compared to the regular radiators. In a real case, photoelectric absorption in the radiator leads to attenuation of the low energy part of the signal. Fig. 2 shows only the emitted radiation, but given the comparable total thickness of the regular and compound radiators in g/cm$^2$, there should be little difference in the X-ray transmission probabilities. 

These examples are specifically for the case of the 
APT gamma-ray/cosmic-ray design. In the case of the FMS accelerator experiment, the total available length along the beam line is comparable to the length of the APT instrument, and the range of desirable $\gamma$ values is the same, so that the examples above apply both to the accelerator and the space applications.

\section{Conclusions}
Transition radiators to date have consisted of either regular radiators with constant values of foil thickness and spacing, or foam or fiber radiators with equivalent average wall thickness  and cell size. In order to extend the range of Lorentz factor sensitivity for future applications and increase the X-ray yield, compound radiators with more complicated sets of radiator parameters may be useful. We have presented a simple derivation of the expected yield for a configuration of $N$ radiator modules each with an arbitrary set of $M$ foils (i.e., $M$ arbitrary choices of foil dimensions $l_i$) by adding the field amplitudes at each interface in phase. We have presented a numerical example showing the produced X-ray energy for a regular foil radiator designed to be compatible with the practical requirements (e.g., maximum length available and desired Lorentz factor range) of 1) a TRD designed to separate 0.5 – 3 TeV protons, kaons, and pions for the FMS experiment at CERN and 2) a TRD designed to measure the spectrum of cosmic-ray nuclei from 500 GeV/nucleon to 20 TeV/nucleon on the proposed APT $\gamma$-ray satellite. A compound radiator provides additional flexibility in the design (at the cost of more complexity in the choice of instrument parameters) and, in the case of this comparison with regular radiators designed for the same FMS and APT experimental conditions, the compound radiator  produces increased signal and a broader range of accessible Lorentz factors.

A future paper will extend the current calculation by using complex values of the index of refraction in order to include the effects of absorption and scttering in the radiator. This will be necessary for a full evaluation of the expected performance of a compound radiator in a real experiment.

\section*{Acknowledgements}
This work has been supported by NASA award  80NSSC19K0625. We appreciate valuable conversations with members of the FMS and APT experiment teams and the ATLAS Transition Radiation group.


\bibliography{NIM_TR_Jan18}

\end{document}